\documentclass[a4paper,11pt]{article}
\usepackage{pos}
\usepackage{bbold}
\usepackage{bm}

\renewcommand{\v}[0]{\bm}
\renewcommand{\tt}{\textit}

\renewcommand{\tt}[1]{\ignorespaces}

\title{Full resummation analysis of jet quenching and tests of the QCD
  Equation of State}
\author*[a]{X. Feal}

\affiliation[a]{Physics Department, Brookhaven National Laboratory\\
  Bldg. 510A, Upton, NY 11973, USA}

\emailAdd{xgarciafe@bnl.gov}

\abstract{Jet quenching has become a fundamental tool to study the hot
  QCD matter produced in heavy ion collisions. While important
  theoretical and experimental advances have been made in the last two
  decades, the extraction of the medium properties and the comparison
  with finite temperature QCD is still particularly worrisome. In this
  work we show that improvements in the calculation of the
  medium-induced gluon spectrum are required for a correct extraction
  of the  parameters without temperature issues. In particular, we
  employ an improved numerical implementation of multiple hard
  scatterings that resums all terms in the opacity expansion beyond
  the Gaussian approximation. We find significant differences in the
  extracted medium parameters when comparing with two of the most used
  approximations in phenomenological analyses to date, the first order
  opacity expansion and the Gaussian approximation. We also make a
  first attempt to compare the extracted medium parameters with
  lattice results.}

\FullConference{%
  HardProbes2020\\
  1-6 June 2020\\
  Austin, Texas}

\begin{document}
\maketitle

The study of the properties of the hot QCD matter constitutes one of
the main goals of the heavy ion program at LHC and RHIC
\cite{akiba2015,Citron,Dainese}. A fundamental tool in these studies
is provided by jet quenching observables. After a typical hard event a
parton is expected to undergo multiple interactions with the hot and
dense medium created in the collision of two heavy ions. Measuring the
imprints of these interactions in the final spectrum of these
particles provides, then, a natural way to explore the properties of
the Quark Gluon Plasma (QGP).

Despite the current wealth of data and the large progress on the
theoretical side over the last decades, the extraction of the QGP
parameters through jet quenching analyses has been subject to some
uncertainties. Current calculations for one of the simplest
observables, the nuclear modification factor $R_{AA}^h$, which
measures the depletion of high $p_\perp$ hadrons in the QGP, have been
in good agreement with the data for a given energy, centrality and
colliding system. However, a common understanding of the RHIC and LHC
data without temperature issues is still missing. Particularly, an
inconsistent deviation in the extracted medium opacity has been
reported by all the phenomenological analyses to date
\cite{jet2014,andres2016,Zakharov:2020whb}, making the QGP produced at
RHIC look apparently $K\simeq$ 1.3-2 times more dense than expected
assuming the LHC temperature scenario. To this extent, it is natural
to wonder whether these discrepancies can be attributed or not to the
required approximations employed so far in the required calculations
to predict the jet quenching phenomena.

The main contribution to jet quenching is the energy-loss of fast
partons in the QGP through multiple gluon emission
\cite{Qin:2007rn}. In a pQCD scenario, partons collide multiple times
with QGP constituents, quantum mechanically rotating its color and
momentum and giving rise thus to radiative corrections depleting the
high energy states before hadronization. The resummation of this
multiple scattering effects in the gluon intensity has been
challenging, and the form of the spectrum after a hard collision
vertex has been known for a while only in some approximations. In the
\textit{single hard} approximation
\cite{gyulassy2000,wiedemann2000b} only the first or first
terms in an expansion in the number of collisions are kept. While this
approximation provides a satisfactory description of the perturbative
cross sections, it does not fully account the whole multiple
scattering in the QGP, lacking then some sensitivity of the emission
to the decoherence phenomena occurring in the soft gluon limit
\cite{landau1953a,landau1953b,migdal1956}. On the other hand, in the
\textit{multiple soft} - or Gaussian - approximation
\cite{zakharov1996,baier1997a,baier1997b,Zakharov:1998sv,salgado2003,armesto2004}
arbitrary large opacities can be handled by relaxing the parton medium
cross sections to Gaussian distributions. While this approximation
seems good for the account of multiple soft scatterings, it misses the
typically large tails $1/q^4$ of the real cross sections, signaling
collisions with point-like scatterers in the plasma, and thus lead to
a substantial underestimation of the large $1/\omega$ tails of the
gluon spectrum.

These caveats and the issues in the extraction of the QGP parameters
have recently triggered more precise calculations of the medium
induced gluon radiation
\cite{apolinario2014,sievert2018,feal2018a,Sievert:2019cwq,Mehtar-Tani:2019tvy,Andres:2020vxs,Barata:2020sav}. We
propose here to determine if the absence of a full resummation of the
\textit{multiple hard} scattering has been consistently re-absorbed as
a systematic error in the extracted parameters. To that end, we
compare the extracted QGP opacity using either the first order
perturbative result or including all the neglected terms in this
expansion. We are able to show that this improvement is indeed
required for a correct extraction of the QGP parameters without
temperature issues. By the same token, we will be able to compare our
results with the lattice predictions, opening up an additional handle
to study the QCD Equation of State.

The intensity of soft gluons emitted in an in-medium path $l$ with
energy between $\omega$ and $\omega+d\omega$, per unit of time and per
unit of medium transverse size, up to all orders in the opacity, is
given by
\begin{align}
\omega \frac{dI}{d\omega} = \alpha_s C_R&
\int \frac{d\Omega_n}{(2\pi)^2}\left(\prod_{k=0}^{n-1}\int 
\frac{d^3\v{k}_{k}}{(2\pi)^3}\phi(q_k,
\delta s_k) \right)J_T^2(k),
\label{squared_emission_amplitude}
\end{align}
where $\alpha_s$ is the coupling constant and $C_R=4/3 (3)$ is the
color averaged squared vertex for a gluon radiated off a quark (gluon)
leg. The parton momentum distributions in the QGP are encoded as color
and spatial averages of squared S-matrices in a classical background
gauge field of tagged partons, and given by
\begin{align}
  \phi(q,s) = 2\pi\delta(q^0) \int d^2\v{x}e^{i\v{q}\cdot\v{x}}\exp\int^s_0 dz
  \rho(z)\bigg[\sigma(\v{x}) -\sigma(\v{0})\bigg],
  \label{multiple_elastic_distribution}
\end{align}
where $q$ is the momentum exchange of the gluon in a path of length
$s$, $\rho(z)$ is the local number density of the medium at a depth
$z$ and $\sigma(\v{x})$ is the color averaged Fourier transform of the
single elastic cross section for the leading order scattering
amplitude
$F_{el}^{(1)}=-i4\pi\alpha_st_{\alpha}^At_{\alpha}^R/(\v{q}^2+\mu_d^2(T))$
with $t_{\alpha}^{A,R}$ the SU(3) matrices in the adjoint
representation $A$ and the target parton representation $R$. The total
squared emission current contains only the subset of gluon
bremsstrahlung diagrams contributing to the intensity after a hard
production vertex
\begin{align}
J_T^2(k)=\bigg|J^n+\sum_{l=1}^{n-1}
J^l\bigg|^2-\bigg|J^n\bigg|^2,\medspace\medspace\medspace  J^l_i= \epsilon_{ijk}\frac{ k_j^lp_k}{k_\mu^l p^\mu}
  \left(e^{i\varphi_{l+1}^n}-e^{i\varphi_{l}^n}\right),
  \medspace\medspace\medspace
  J^n_i= \epsilon_{ijk}\frac{ k_j^np_k}{k_\mu^n p^\mu},
\label{total_current}
\end{align}
where in the Coulomb gauge $J_0^l=J_0^n=0$ and $\epsilon_{ijk}$ is the Levi-Civita
symbol. The hard particle 4-momentum $p=(p_0,\v{p})$ can be left fixed, in the
soft gluon limit, in the particle initial direction after the
production vertex. The phase difference $\varphi^{a}_{b}$ records the
4-current incoherence introduced between multiple interactions at $a$
and $b$ and modulates the non-Abelian LPM effect. It reads
\begin{align}
  \varphi^n_l=\frac{1}{p_0}\sum^{n-1}_{i=l}\delta s_i k^i_\mu p^\mu,
  \medspace\medspace\medspace
  \delta s_i = s_{i+1}-s_i,
\end{align}
where $s_i$ is the interaction point in the path and $p_0$ the hard
particle energy.

To make connection with previous works, intensity in
Eq. \eqref{squared_emission_amplitude} resums at the amplitude level
all the multiple scattering up to arbitrary and finite opacities,
considering the full kinematics of the interactions, and admitting
arbitrary forms of the perturbative cross sections and arbitrary
profiles for static or expanding media
\cite{baier1998,salgado2002}. In the collinear limit $\omega \gg
k_\perp$ well known results in the light cone are found
\cite{feal2018b}. In particular, path integrals matching the BDMPS-ASW
\cite{baier1997a,salgado2003} result in the Fokker-Planck
approximation for \eqref{multiple_elastic_distribution}, or the more
recent numerical implementation of the BDMPS-ASW expression for
multiple hard scatterings \cite{Andres:2020vxs}. On the other hand,
the perturbative expansion reproduces the GLV series
\cite{gyulassy2000,wiedemann2000b}. In this collinear approximation,
resummations for semi-infinite \cite{zakharov1996,arnold2002} and
finite media \cite{CaronHuot:2010bp} have been known for a while with
an approximate integration of the angular distribution of gluons. In
the soft limit the distribution of multiple gluon emission is given by
a classical Poisson process. Then the probability $P$ of a parton of
initial energy $p_\perp$ of losing an energy $\epsilon$ is given by
\begin{align}
  P(\epsilon,p_t)=\mathcal{N}\sum_{n=0}^\infty \frac{1}{n!}
  \delta\bigg(\epsilon-\sum_{i=1}^n\omega_i\bigg)\prod_{i=1}^n
  I\bigg(p_t-\sum_{j=1}^{i-1}\omega_j\bigg).
  \label{quenching_weight}
\end{align}
where $N$ is a normalization constant. The quenched cross section of
observing an hadron of energy $p_\perp$ is then given by
\cite{baier2001a,miller2007}
\begin{align}
  \frac{d\sigma_{AA}^h(p_t)}{dydp_t}\simeq T_{AA}\sum_{i}\int^{\infty}_0 d\epsilon
  \frac{d\sigma_{pp}^{i,h}(p_t+\epsilon)}{dy dp_t}P_i(\epsilon,p_t+\epsilon).
  \label{hadron_cross_section}
\end{align}
where $i$ is the parton species, a light quark representative at high
$p_\perp$ and a gluon at low $p_\perp$, and $T_{AA}$ the nuclear
overlap function. In \eqref{hadron_cross_section} the fragmentation
function effects have been averaged at a typical fragmentation
ratio. Finally the nuclear modification factor is defined as
\begin{align}
R_{AA}^{h}\equiv  \frac{1}{\langle T_{AA}\rangle
}\frac{d\sigma_{AA}^{h}(p_t)}{dydp_t}\bigg/
\frac{d\sigma_{pp}^{h}(p_t)}{dydp_t}.
\label{nuclear_modification_factor}
\end{align}

To test the sensitivity of the nuclear modification factor
\eqref{nuclear_modification_factor} to the spectrum resummation, a QGP
is implemented with Bjorken, longitudinal expansion with equilibrium
temperature $T_0$ set by the energy density measurements and the
scaling $\epsilon\tau_0\propto T_0^3$, where $\tau_0\propto 1/T_0$ is
the thermalization time \cite{strickland2018}. The relation
$\epsilon\tau_0 =(8.85\pm 0.44)\times
\left(\sqrt{s_{nn}}\right)^{0.33\pm 0.02}$, found for the most central
classes between $\sqrt{s_{nn}}$=27GeV-2.76TeV
\cite{phenix2016,alice2016b} is instead used when data has not been
made available. Then, a reference temperature is set for the most
central PbPb collisions at $\sqrt{s_{nn}}$=2.76 TeV to the value
$T_0=470$ MeV, following \cite{kolb2003}. Once this temperature is
set, all the parameters in the analysis are being fixed except for the
QGP number density, which is the only unknown parameter given by the
fit to the $R_{AA}^h$ data. The Debye screening mass $\mu_d(T)$,
characterizing collisions with point-like scatterers in the QGP
\cite{DEramo:2012uzl}, and the strong coupling constant $\alpha_s(T)$,
are given at leading order by $\mu_d^2(T)=4\pi\alpha_s(T)(1+N_f/6)T^2$
\cite{braaten1990,rebhan1993} and
$\alpha_s(q)=1/(b_0\ln(q^2/\Lambda^2))$, respectively, with $b_0$ the
1-loop $\beta$-function coefficient for $N_f=2+1$ active flavors, and
the momentum scales are set as $q=2\pi T$ and $\Lambda_s=247$ MeV to
match the world averaged data in \cite{pdb2018}. This setup reproduces
well the collected data up to the coldest system analyzed in this
work, Cu-Cu collisions at $\sqrt{s_{nn}}$=200GeV at centrality
$30\%-40\%$, and the color screening reproduces very well the pure
gauge ($N_f=0$) lattice computations, and falls slightly below of the
$N_f=2$ flavor result in the staggered quark action
\cite{zantow2005}. The QGP lifetime is set with the relation
$\tau_f=(0.87\pm 0.01)\times(dN_{ch}/d\eta)^{1/3}$ fm/c found for the
available data \cite{star2005b,alice2015} on Bose-Einstein pion
correlations\cite{makhlin1988,retiere2004} , with kinetic freeze-out
set to $T_f=120$ MeV. For further details we refer to \cite{feal2019}.

\begin{figure}[ht]
  \centering \includegraphics[scale=0.55]{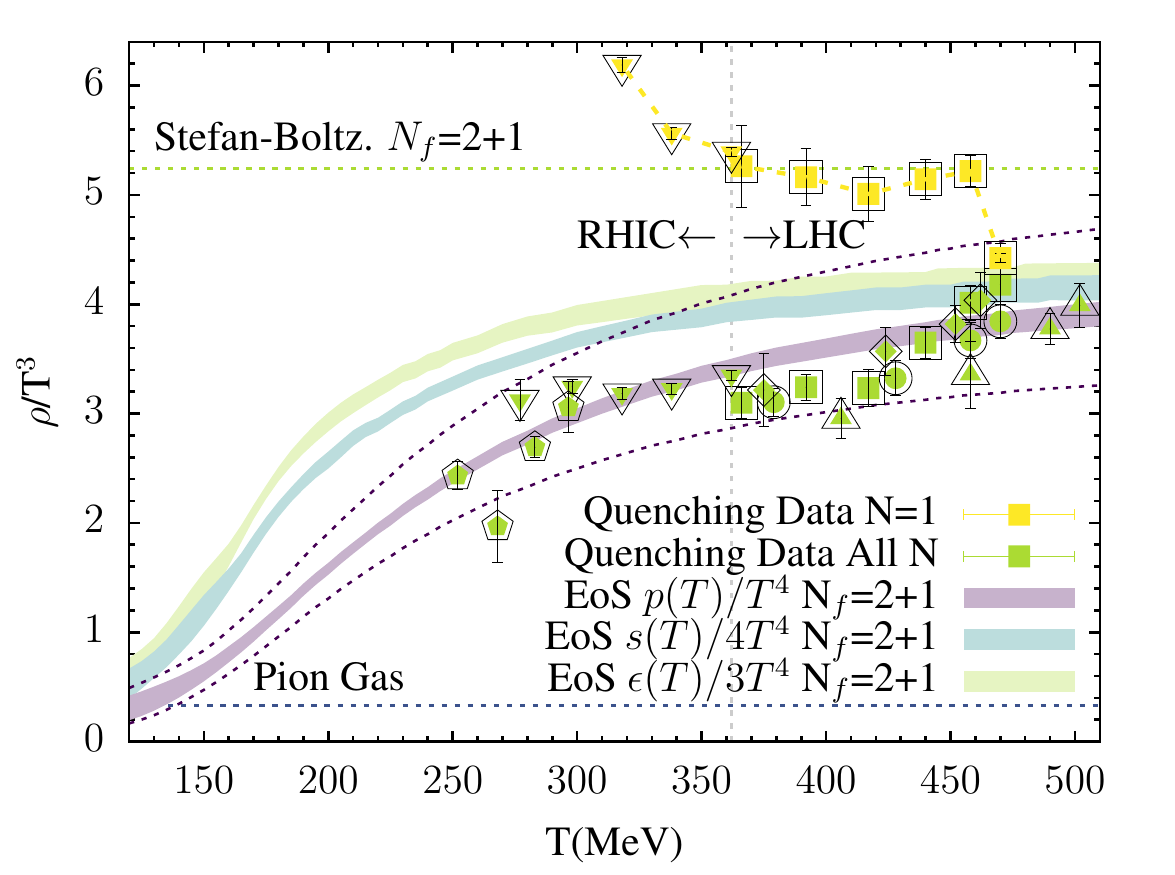}
  \includegraphics[scale=0.55]{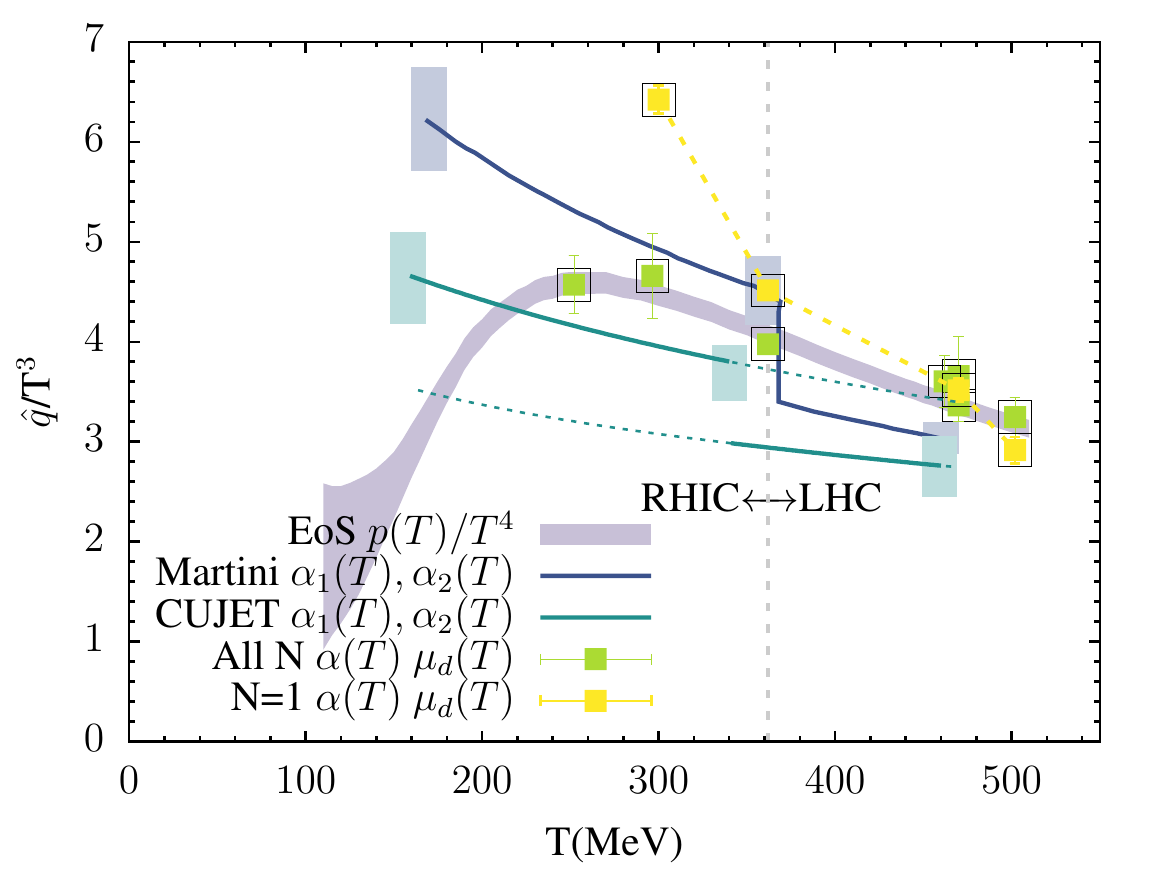}
  \caption{[Left] QGP density extracted from an All $N$ analysis
    (purple symbols ) and a first order $N=1$ analysis (green symbols)
    of the $R_{AA}^h$ collected data on collisions of \textit{CuCu}
    (pentagons) and \textit{AuAu} (down triangles) at
    $\sqrt{s_{nn}}=$200 GeV, \textit{PbPb} at $\sqrt{s_{nn}}$=2.76 TeV
    (squares and circles), \textit{PbPb} at $\sqrt{s_{nn}}$=5.02 TeV
    (up triangles) and \textit{XeXe} at $\sqrt{s_{nn}}$=5.44 TeV
    (diamonds) from PHENIX, ALICE and CMS Collaborations, compared to
    lattice results of the Equation of State by the Wuppertal
    collaboration \cite{wuppertal2016}. [Right] QGP transport
    parameter $\hat{q}$ extracted from an all order analysis (green
    squares) or a fist order order analysis (yellow squares) of the
    $R_{AA}^h$ data. Also shown is the $\hat{q}$ assuming $\rho=p/T^4$
    from lattice predictions of the QCD Equation of State
    \cite{wuppertal2016} (green band), and the CUJET (blue) and
    MARTINI (purple) puzzles \cite{jet2014}.}
  \label{figure_1}
\end{figure}

An analysis of existing data on jet quenching, including Cu-Cu and
Au-Au data at 200 GeV \cite{phenix2008a,phenix2008b}, Pb-Pb at 2.76
TeV \cite{cms2012,alice2013a,alice2011}, Pb-Pb at 5.02 TeV
\cite{cms2017,alice2016c}, and Xe-Xe data at 5.44 TeV \cite{cms2018}
is made using the full order resummation as well as only the first
term in the opacity expansion. For each case a fit is made to the
nuclear modification factor $R_{AA}^h$ using $\rho(T_0)$ as the single
unknown parameter.  Results are collected in Fig.\ref{figure_1}. The
found opacity using the full resummation of the spectrum roughly
scales as $T_0^3$ with a slow decrease along the phase
crossover. Unlikely, when only the first term in the perturbative
expansion is kept, substantial deviations from the $T_0^3$ behavior
are found as the temperature of the QGP decreases, energy to energy,
centrality to centrality. A deviation factor of $K\simeq 1.22$, or of
$K\simeq 1.29$ in terms of $\hat{q}$, the QGP transport parameter, is found
when going from the most central Pb-Pb collisions at
$\sqrt{s_{nn}}=$2.76 TeV in the LHC to the most central Au-Au
collisions at $\sqrt{s_{nn}}=$200 GeV at RHIC. Results for $\hat{q}$
are shown in Fig.\ref{figure_1}. This factor is consistent with the
value $K\simeq 1.3$ observed in the Jet Collaboration analysis
\cite{jet2014} using the $N=1$ approximation. Larger factors $K\simeq 2$
have been found using the Gaussian approximation \cite{andres2016}. We
thus conclude that for a description of the jet quenching data without
temperature issues a full resummation of the medium induced gluon
radiation is required. In addition, since the scalings seem to be
consistent with the high temperature limit extrapolations, we compare
our results for $\rho(T_0)$ with the lattice predictions for the
entropy $s/4T^3$, pressure $p/T^4$ and energy density
$\epsilon/3T^4$. These predictions are also shown in
Fig. \ref{figure_1}. Our result for the number density seems to agree
rather well with the lattice predictions for the pressure, although
under our current uncertainties - further discussed in \cite{feal2019}
- it may be also consistent with either the entropy or energy density
relations. This preliminary study may pave the path for more precise
analyses, providing and additional handle to study the QCD
Equation of State.

\acknowledgments

We thank the very valuable contributions of R.A. Vazquez and
C.A. Salgado to this work. This work has been funded by Ministerio de
Ciencia e Innovaci\'on of Spain under project FPA2017-83814-P; Unidad
de Excelencia Mar\'\i a de Maetzu under project MDM-2016-0692;
ERC-2018-ADG-835105 YoctoLHC; and Xunta de Galicia (Conseller\'\i a de
Educaci\'on) and FEDER. X.F. is supported by grant ED481B-2019-040
(Xunta de Galicia) and the Fulbright Visiting Scholar fellowship.

\end{document}